\documentstyle[epsf]{mn}

\title{Detection of the optical counterpart of the proposed double
degenerate polar RX J1914+24}

\author[G. Ramsay et al.]{Gavin Ramsay$^{1}$,
Mark Cropper$^{1}$, Kinwah Wu$^{2}$, K. O. Mason$^{1}$, Pasi
Hakala$^{3}$\\
$^{1}$Mullard Space Science Laboratory, University College London, 
Holmbury St. Mary, Dorking, Surrey RH5~6NT\\
$^{2}$Research Centre for Theoretical Astrophysics, School of Physics,
University of Sydney, NSW 2006, Australia\\
$^{3}$Observatory and Astrophysics Lab, FIN-00014, Univ Helsinki,
Helsinki, Finland}

\date{Received: }

\begin{document}
\def\Mdot{\hbox{$\dot M$}}
\def\Msun{\hbox{$M_\odot$}}
\outer\def\gtae {$\buildrel {\lower3pt\hbox{$>$}} \over 
{\lower2pt\hbox{$\sim$}} $}
\outer\def\ltae {$\buildrel {\lower3pt\hbox{$<$}} \over 
{\lower2pt\hbox{$\sim$}} $}

\maketitle

\begin{abstract}

We have detected the optical counterpart of the proposed double
degenerate polar RX J1914+24. The $I$ band light curve is modulated on
the 9.5 min period seen in X-rays. There is no evidence for any other
periods. No significant modulation is seen in $J$. The infrared
colours of RX J1914+24 are not consistent with a main sequence dwarf
secondary star. Our {\sl ASCA} spectrum of RX J1914+24 is typical of a
heavily absorbed polar and our {\sl ASCA} light curve also shows only
the 9.5 min period. We find that the folded $I$ band and X-ray light
curves are out of phase. We attribute the $I$ band flux to the
irradiated face of the donor star. The long term X-ray light curve
shows a variation in the observed flux of up to an order of
magnitude. These observations strengthen the view that RX J1914+24 is
indeed the first double degenerate polar to be detected. In this light,
we discuss the synchronising mechanisms in such a close binary and
other system parameters.

\end{abstract}


\section{Introduction}

Cropper et al.\ (1998) reported X-ray observations of the cataclysmic
variable (CV) RX J1914+24 which was discovered by Motch et al.\ (1996)
in the {\sl ROSAT} All-Sky Survey. Motch et al also found that the
X-ray flux was modulated at 9.5 min (= 569 sec). This period is
characteristic of Intermediate polars (IPs: the non-synchronous
magnetic CVs, Patterson 1994), but the shape of the folded light curve
showed no X-ray flux for approximately half the 9.5 min period. This
property is difficult to reconcile with our current understanding of
IPs. Given the absence of any other periods, Cropper et al.\ (1998)
suggested that RX J1914+24 was a polar, or AM Her system -- the
synchronous magnetic CVs (Cropper 1990, Beuermann \& Burwitz
1995). This would imply that the binary orbital period is 9.5 min.

An orbital period of 9.5 min would be the shortest period of any known
binary star. It would also exclude the secondary from being a
main-sequence star (the shortest period allowed is $\sim$ 80~min). For
a non-degenerate He burning secondary, the minimum orbital period is
$\sim$ 13~min (Iben \& Tutukov 1991). However, a degenerate He
secondary can have a period in the range $\sim$ 6 -- 50 min (Tutukov
\& Yungelson 1996). Confirmation that RX J1914+24 is a
double-degenerate polar would have important implications for how
magnetic binaries evolve. It would also be the first known magnetic
system in which the accretion flow was predominately He dominated.

Comparison between the X-ray position of RX J1914+24 (determined using
the High Resolution Imager, HRI, on {\sl ROSAT}) and the optical image
of Motch et al (1996) showed that star `H' was the most likely optical
counterpart (Cropper et al 1998). This star is heavily reddened
($A_{V}\sim5.6$: Motch et al 1996). To detect the optical counterpart
of RX J1914+24 and search for periods other than the 9.5 min period
seen in X-rays we have obtained images of the field of RX J1914+24 in
the $I$ band and at IR wavelengths, the rationale being that if no
other periods were seen then this would strengthen the conclusion of
Cropper et al. (1998) that RX J1914+24 is a double degenerate
polar. We also report another set of X-ray data obtained using {\sl
ROSAT} and one obtained using {\sl ASCA}.

\section{X-ray observations}

\subsection{{\sl ROSAT} Data}
\label{rosat}

\begin{figure}
\begin{center}
\setlength{\unitlength}{1cm}
\begin{picture}(8,14)
\put(-1.5,-1){\includegraphics{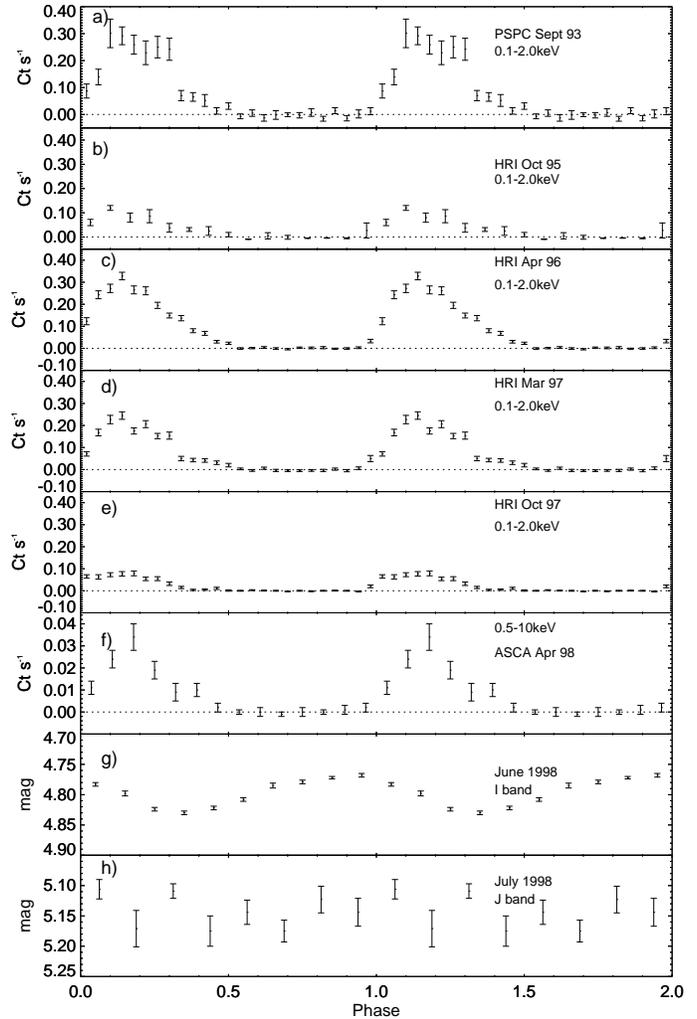}}
\end{picture}
\end{center}
\caption{Panels a)--e): the {\sl ROSAT} data from different epochs,
Panel f): the ASCA data, Panel g) the I band data obtained using NOT and
Panel h): J band data obtained using UKIRT.}
\label{fold} 
\end{figure}

Cropper et al (1998) presented X-ray observations of RX J1914+24 taken
using {\sl ROSAT} (0.1--2.0keV) at 3 distinct epochs. While the shape
of the folded light curve was similar at each epoch, the peak
intensity varies over time. To monitor the shape and intensity of the
X-ray light curve we obtained a further set of data using the {\sl
ROSAT} HRI in Oct 1997 with a the total exposure of 21.5ksec. We
extracted a background subtracted light curve in the same way as
Cropper et al.  (1998) and show it along with the other {\sl ROSAT}
data in Fig.\ref{fold}. We also show the folded light curve of the HRI
data taken in Oct 1995 which was not shown in Cropper et al. (1998) as
the exposure time was only 2.4ksec. These data show a range of maximum
brightness: using a spectrum similar to that found in \S \ref{asca}
for the {\sl ASCA} data, the system was a factor of $\sim$3 fainter in
Sept 1993 compared to that in Apr 1996.

The data shown in Fig.\ref{fold} were folded on the linear ephemeris
obtained by fitting the start of the rise from zero flux. From 4
epochs we obtained 108 timings. This gave a best fit ephemeris:

\vspace{1mm}
T$_{o}$= HJD 2449258.03941(6) + 0.0065902334(4)
\vspace{1mm}

\noindent The numbers in brackets give the standard error on the last
digit. This allows us to phase all our data from Sept 1993 to July
1998 within 0.02 cycles.

\subsection{{\sl ASCA} observations}
\label{asca}

We obtained a 20ksec observation of RX J1914+24 in 1998 April 9--10
using {\sl ASCA} ($\sim$0.5--10keV). The data were extracted by
applying the standard data selection criteria. The background was
generated using a concentric annulus around the source. RX J1914+24
was detected only in the SIS detector and only below 1keV. The light
curve from the SIS0 detectors (the peak flux of the folded SIS1 data is
approximately half that of SIS0) were analysed using a Discrete
Fourier Transform code (DFT; Deeming 1975, Kurtz 1985). The amplitude
spectrum is shown in Fig. 2: the most prominent amplitude peak
corresponds to the 9.5 min period seen in the {\sl ROSAT} data. The
amplitude spectrum pre-whitened by the 9.5 min period (and harmonic)
is also shown in Fig.2. There is no evidence for a significant
modulation at any other period.

The light curve derived from the SIS0 data was folded on the ephemeris
determined in \S \ref{rosat}.  We show the folded light curve in
Fig. \ref{fold}.  The shape of the light curve during the bright
phase is more peaked than the {\sl ROSAT} HRI data taken in Oct 1997
but similar to that taken in Apr 1996. Using the spectral parameters
found below we find that RX J1914+24 was fainter in X-rays by a factor
of $\sim$3 at the epoch of the {\sl ASCA} observation compared with
Oct 1997.

\begin{figure*}
\begin{center}
\setlength{\unitlength}{1cm}
\begin{picture}(8,10)
\put(-5,-28.5){\includegraphics{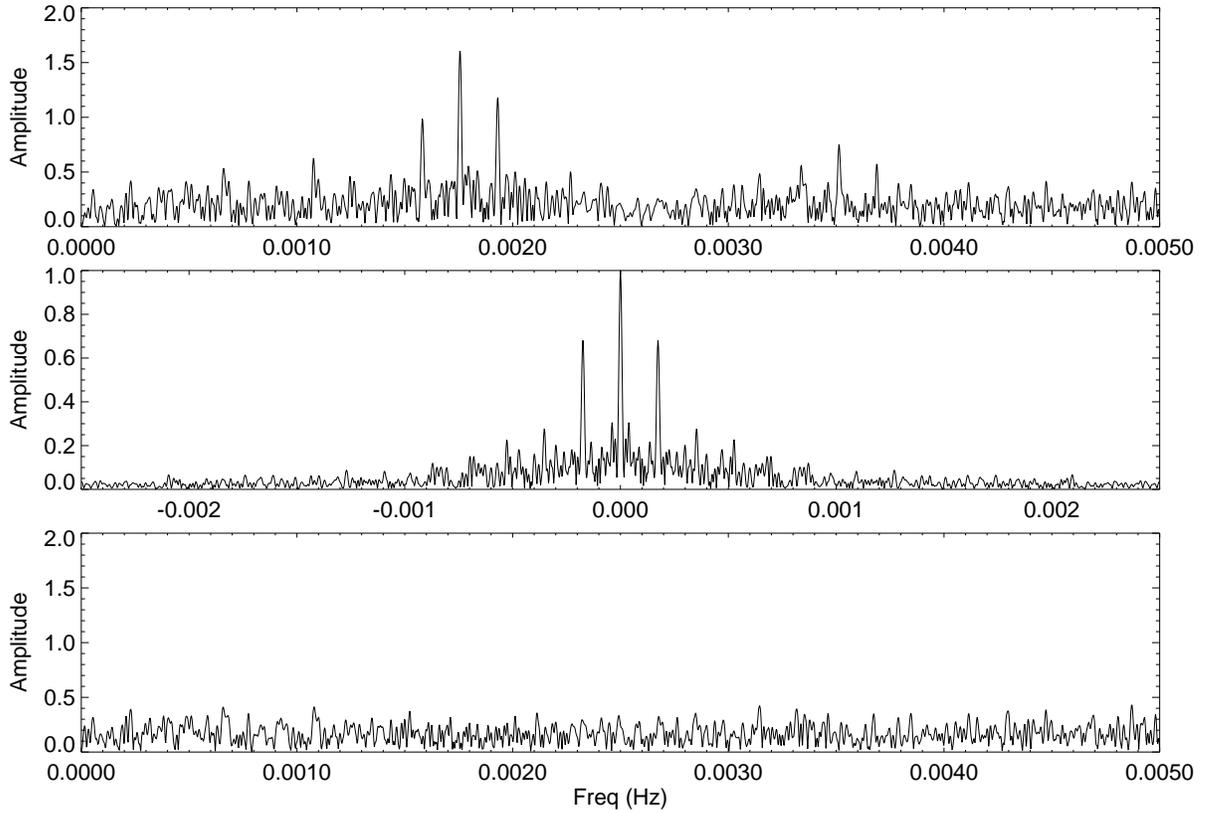}}
\end{picture}
\end{center}
\caption{The amplitude spectrum (top) of the {\sl ASCA} SIS0 data, the
window function (middle) and the amplitude spectrum pre-whitened by the
569 sec period (lower).}
\label{asca_amp} 
\end{figure*}

\begin{figure}
\begin{center}
\setlength{\unitlength}{1cm}
\begin{picture}(8,7)
%\put(-3.5,-5.5){\special{psfile=../mcv2/rx1914_Fig..eps}}
\put(-9.,-0.7){\includegraphics{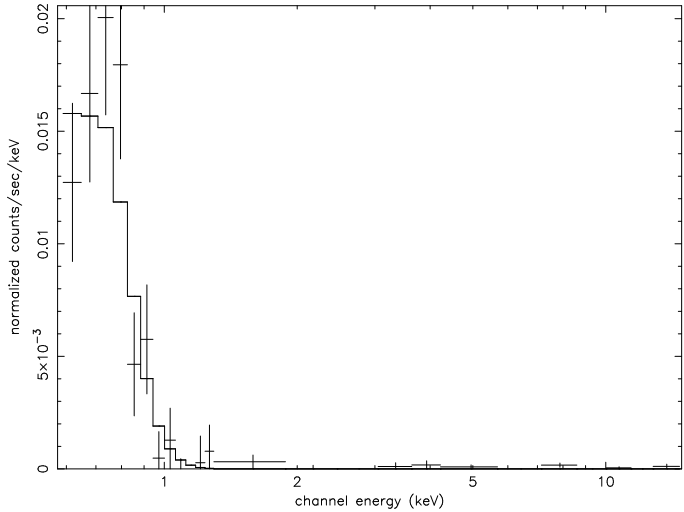}}
\end{picture}
\end{center}
\caption{The integrated {\sl ASCA} SIS-0 spectrum together with the best fitting
absorbed blackbody model.}
\label{asca_spec} 
\end{figure}

The {\sl ASCA} spectrum (Fig. \ref{asca_spec}) could be fitted
reasonably well (the best fit gave $\chi^2$ = 1.31, 20 degrees of
freedom), with a low temperature ($kT\sim$40eV) blackbody plus
interstellar absorption (N$_{H}\sim1\times10^{22}$ cm$^{-2}$). The
temperature and absorption were not well constrained. The upper limit
(90 percent confidence) to a thermal bremsstrahlung component (using
an assumed temperature of 10keV) corresponds to a 2--10 keV flux of
$7.8\times10^{-14}$ erg cm$^{-2}$ s$^{-1}$ ($9\times10^{28}$ erg
s$^{-1}$ at 100pc). The bolometric luminosity of the blackbody
component is $\sim1\times10^{36}$ erg s$^{-1}$ at 100pc. This agrees
very well with the estimate derived from the {\sl ROSAT} spectrum
(Cropper et al 1998). The low flux from the bremsstrahlung component
and the very large ratio between the soft blackbody component and the
hard bremsstrahlung component is typical of polars (eg Ramsay et al
1994). X-ray spectra of IPs typically have a strong medium energy
component and rarely show a strong soft component.

\section{Optical-IR Observations}

\subsection{$I$ band polarimetry}

\begin{figure}
\begin{center}
\setlength{\unitlength}{1cm}
\begin{picture}(7,6)
\put(13.5,14.5){\includegraphics{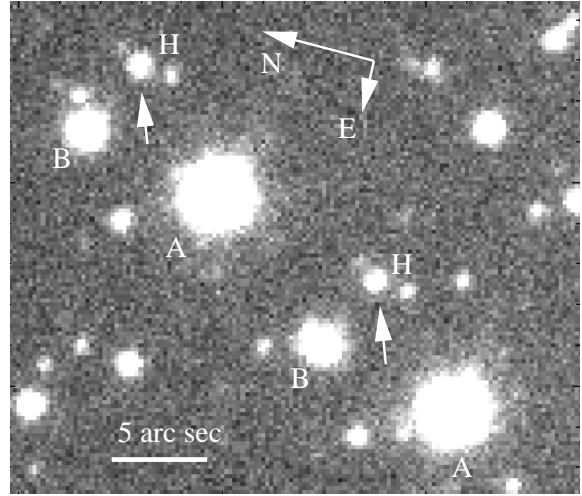}}
\end{picture}
\end{center}
\caption{An $I$ band image of the field of RX J1914+24 taken in 
   June 1998 using NOT. Two images of the each object in the field are 
   recorded -- one from each polarised ray. Stars `A' and `B' in the 
   finding chart of Motch et al.\ (1996) are shown. RX J1914+24 is 
   marked by an arrow. Star `H' of Motch et al.\ is thus a blend of more 
   than 1 star (including RX J1914+24).}
\label{iband} 
\end{figure}

Observations were obtained on the nights 1998 June 25 -- 27 using the
2.5-m Nordic Optical Telescope (NOT) on La Palma and the Andalucia
Faint Object Spectrograph and Camera (ALFOSC) used in its imaging
mode. On the first night, data were obtained over 7hrs 26min while
this figure was 4hrs 19min and 7hrs 55mins on the second and third
nights respectively. The conditions were good and the seeing was
typically 1$^{''}$ or better. To obtain circular polarimetry data a
1/4 waveplate was inserted into the optical path: this split the light
into o and e rays on the Loral 2k$\times$2k CCD so that 2 images of
each star were recorded. The instrument was orientated in such a way
that images of field stars did not overlap with any of the stars of
interest. The images were bias subtracted and flat fielded. A typical
image is shown in Fig.\ref{iband}. In the $I$ band finding chart of
Motch et al.\ (1996) star `H' is shown as a single star: in our $I$
band images (which were taken under good seeing) we find that star `H'
is made up of more than one star. This is consistent with our $K$ band
image taken using UKIRT in service time in October 1997.

\begin{figure*}
\begin{center}
\setlength{\unitlength}{1cm}
\begin{picture}(13,10)
\put(-3,-28.5)
 {\includegraphics{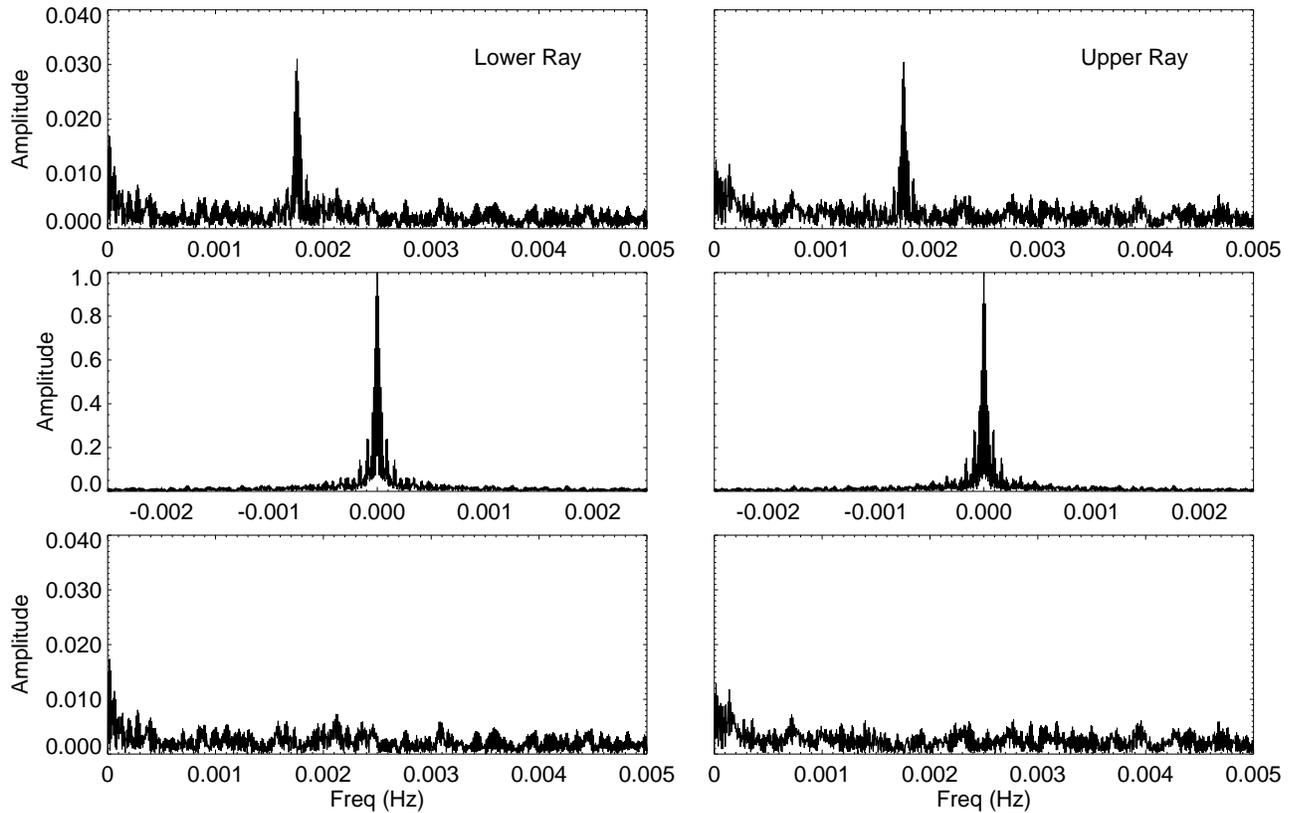}}
\end{picture}
\end{center}
\caption{The amplitude spectra from the lower and upper polarised
images of the brightest component of star `H' are shown in the top
panel. The prominent amplitude peak corresponds to the 569-sec 
(9.5-min) period seen in X-rays. The middle panel shows their window 
functions and the
lower panel shows their amplitude spectra pre-whitened by the 569-sec
period.}
\label{ipower} 
\end{figure*}

Profile fitting photometry was carried out on the stars in the field
using {\tt DAOPHOT} (Stetson 1992). Since the profiles of the stellar
images in the o and e rays differed, two profiles had to be
constructed -- one for each ray. We obtained differential photometry
between Star `A' and the stars in the immediate vicinity of star `H'
and Fourier Transformed their light curves. The brightest component
making up star `H' was found to show a significant amplitude peak at
9.5 min in both the `upper' and `lower' polarised beams (see
Fig. \ref{ipower}). This period corresponds to the period found in
X-rays: this star is therefore the optical counterpart of RX J1914+24
(arrowed in Fig.\ref{iband}). Neither star `B' nor the fainter
component of star `H' were found to show any significant
modulation. Whilst there are minor amplitude peaks at longer periods
in both the `upper' and `lower' amplitude spectra, there is no peak
which is seen in both spectra: therefore we conclude that these
peaks are not significant. 

We folded the $I$ band optical data of RX J1914+24 on the ephemeris in
\S \ref{rosat}. These data are also shown in Fig.\ref{fold}. The
peak-to-peak amplitude is $\sim$ 0.07 mag and the mean magnitude is $I
\sim$ 18.2 assuming the magnitude of star `A' is $I \sim$ 13.4 (Motch
et al.\ 1996). This contrasts with the approximate magnitude of star
`H' as seen by Motch et al.\ in June 1993 of 16.6 mag. Even taking
into account that star `H' is consists of more than 1 component, (RX
J1914+24 being 18.2 mag, the next brightest $\sim$ 19.6), RX J1914+24
was over 1~mag fainter in $I$ in June 1998 compared to June 1993.

If there is a second period hidden the $I$ band data then its
amplitude must be very small. From the noise level in the amplitude
spectrum (Fig. \ref{ipower}), any second period shorter than $\sim$5
hrs must have a peak-to-peak amplitude less than 0.02 mag. This is
much less than the variation generally seen in $I$ band photometry of
other magnetic CVs.  For example in the case of the IP YY Dra, the
orbital period of 3.97 hrs is clearly seen as a modulation with an
amplitude of 0.12 mag in the $I$ band (Haswell et al 1997). Since YY
Dra is expected to have an accretion disc, the modulation of the
secondary star will be diluted by the presence of a disc. In the case
of discless systems such as RX J1914+24 (cf \S \ref{magfield}), the
amplitude is expected to be greater. For instance $I$ band photometry
of the polar V895 Cen in a low state (where there is no additional
flux from cyclotron emission) shows a peak to peak amplitude of
$\sim$0.4 (Stobie et al 1996). The lack of such a pronounced
modulation in RX J1914+24 is further evidence that the secondary star
in this system is not a main sequence star.

\begin{figure}
\begin{center}
\setlength{\unitlength}{1cm}
\begin{picture}(6,9)
\put(-2.5,-6){\includegraphics{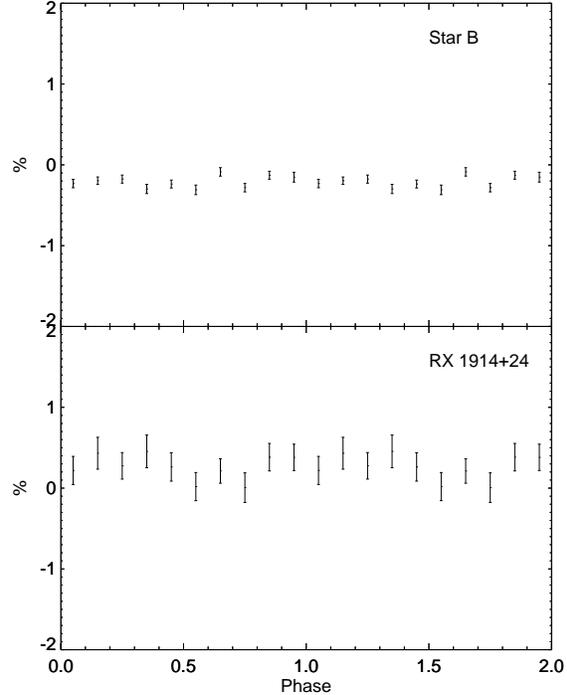}}
\end{picture}
\end{center}
\caption{The circular polarisation data for star `B' in the finding
   chart of Motch et al.\ (1996) and RX J1914+24. The unfolded data 
   were corrected by setting the circular polarisation for star `A' to
   zero. The corrected data were then folded on the ephemeris derived 
   in \S \ref{rosat}.}
\label{pol} 
\end{figure}

The 1/4 waveplate and calcite block in the optical path allowed us to
search for circular polarisation in RX J1914+24. The circular
polarisation was derived for stars `A' and `B' in the finding chart of
Motch et al. (1996) and also RX J1914+24. The instrumental
polarisation was found to depend on the position of the stellar image
on the CCD. We made a first order correction for instrumental
polarisation based on the assumption that star `A' has no intrinsic
circular polarisation. The corrected data of star `B' and RX J1914+24
were then folded on the ephemeris derived in \S \ref{rosat} and are
shown in Fig.  \ref{pol}. Star `B' shows a small net negative circular
polarisation indicating that there is probably some small residual
effects of the polarisation varying on the field position. In the case
of RX J1914+24 there is a small positive ($\sim$ 0.3 per cent)
residual polarisation, although there is no significant modulation
over the 9.5 min period. We conclude that there is no intrinsic
circularly polarised emission from RX J1914+24 in the $I$ band.

\subsection{Infrared Photometry}

RX J1914+24 was observed using UKIRT on Mauna Kea on the nights 1998
July 2 -- 3 mainly in the $J$ band, but images were also obtained in
$H$ and $K$. On the first night conditions were good but in the second
half of the second night conditions were poor and unusable. Images
were obtained using IRCAM3 operating in ND\_STARE mode. In the $J$
band, a 12 sec exposure was made then the telescope was shifted: this
allowed us to make a flat field. Each resulting image was the result
of 6 exposures.

\begin{figure*}
\begin{center}
\setlength{\unitlength}{1cm}
\begin{picture}(13,8)
\put(-3,-32){\includegraphics{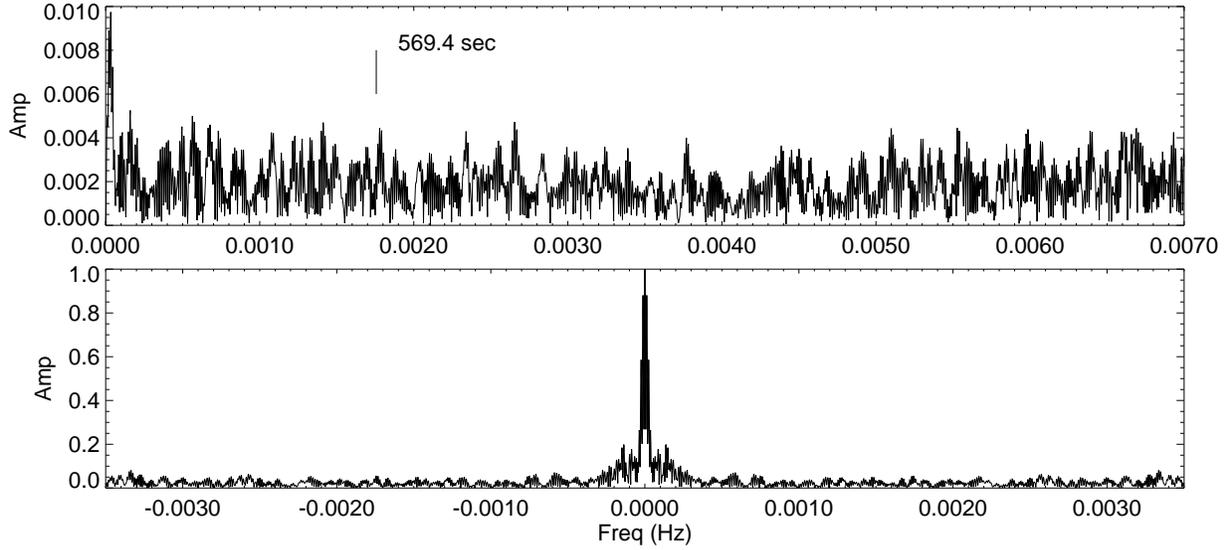}}
\end{picture}
\end{center}
\caption{The amplitude and window spectrum for the $J$ band light
curve obtained at UKIRT. The 569.4 sec marker shows the period of RX
J1914+24 found in X-rays.}
\label{jpower} 
\end{figure*}

Profile fitting was performed on the stars in each image using {\tt
DAOPHOT} (Stetson 1992). As with the NOT data, differential photometry
was performed on star `B' and the optical counterpart of RX J1914+24
using star `A' as the comparison. The differential light curves were
Fourier Transformed. The amplitude spectrum of RX J1914+24 is shown in
Fig.\ref{jpower}: no significant modulations were seen in the light
curves of either RX J1914+24 or star `B'. In the case of RX J1914+24
we folded the differential photometry on the ephemeris shown in \S
\ref{rosat} as shown in Fig.\ref{fold}. As expected from the amplitude
spectrum there is no evidence for a significant modulation on the
9.5 min period. However, the scatter is much greater than the folded
$I$ band light curve and it is possible that a modulation similar to
that seen in $I$ is hidden in the noise.

What amplitude to we expect to see in the $J$ band in normal CVs with
a main sequence secondary stars? A modulation of $\sim$0.4 mag is seen
in the IP FO Aqr (de Martino et al 1994), while no significant
modulation ($<$0.08 mag) is seen in the IP RX J1238-38 (Buckley et al
1998) which at an orbital period of 84 min is below the period gap.
From the noise level of the amplitude spectrum of the $J$ band
photometry (Fig. \ref{jpower}), the upper limit for a modulation is an
order of magnitude lower than observed in RX J1238-38.  We conclude
that this is a very low level for a main sequence secondary star.

In addition to the $J$ band data we also obtained single images in $H$
and $K$ bands. As before, differential photometry was performed
between the optical counterpart of RX J1914+24 and star `A' and
`B'. These differential magnitudes were placed onto the standard
systems from observations of faint UKIRT standard stars. Table
\ref{irmags} shows the reduced magnitudes. We also show the dereddend
magnitudes and fluxes assuming an extinction of $A_{V}$=5.6 mag
(Cropper et al 1998). This corresponds to $A_{I}$=2.72, $A_{J}$=1.59,
$A_{H}$=0.97 and $A_{K}$=0.63 assuming $R$=3.1 and the observed ratios
in Fitzpatrick (1999). The $K$ observed magnitude of 17.1 compares
with 16.4 when we obtained our service observation in Oct 1997.

\begin{table}
\begin{tabular}{llll}
\hline
Band&Mean     &Dereddened&Dereddened Flux\\
    &Magnitude&Magnitude&erg cm$^{-2}$ s$^{-1}$ \AA$^{-1}$\\
\hline
I&18.2$\pm$0.05&15.5&8.1$\pm0.3\times10^{-16}$\\
J&17.3$\pm$0.10&15.7&1.7$\pm0.2\times10^{-16}$\\
H&17.4$\pm$0.15&16.4&3.3$\pm0.06\times10^{-17}$\\
K&17.1$\pm$0.15&16.5&1.0$\pm0.06\times10^{-17}$\\
\hline
\end{tabular}
\caption{The apparent, together with the 
dereddened magnitudes and dereddened fluxes of
RX J1914+24 assuming an extinction of $A_{V}$ = 5.6 mag on 1998 July
2,3. The error on the mean magnitudes includes the error 
in placing them onto the standard system. The error on the dereddened
flux does not include the uncertainty on the extinction.}
\label{irmags}
\end{table}

\section{The Optical near-IR flux}

\begin{figure*}
\begin{center}
\setlength{\unitlength}{1cm}
\begin{picture}(13,8)
\put(-1,-28.5){\includegraphics{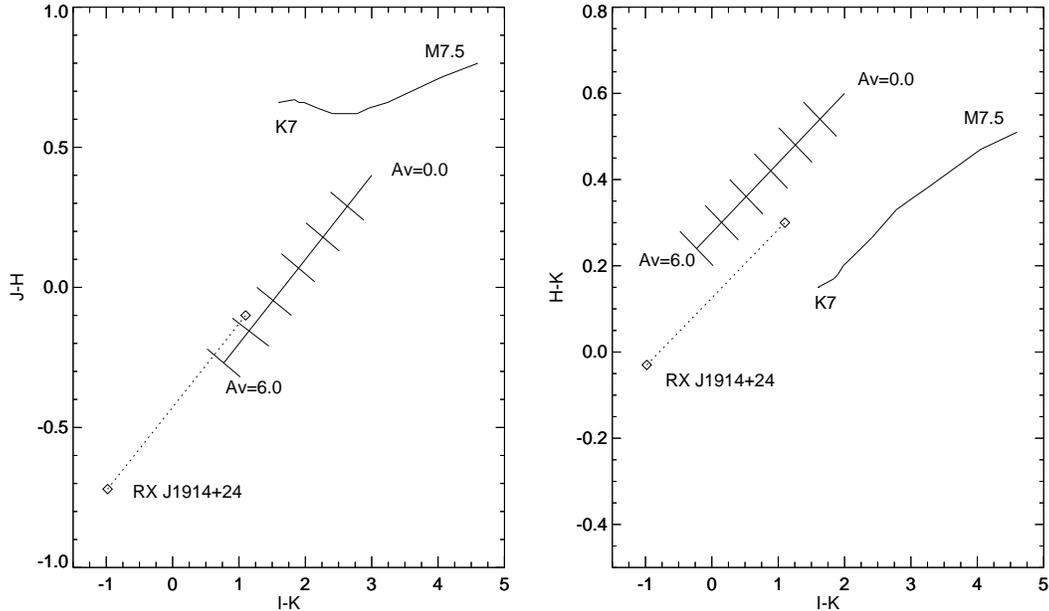}}
\end{picture}
\end{center}
\caption{The position of RX J1914+24 in the observed and dereddened
colours in the ($I-K$), $(J-H)$ and ($I-K$), $(H-K)$ planes. The
positions of old disk M dwarfs with spectral types K7 to M7.5 are also
shown (Bessell 1991). The ticked line shows the reddening direction
for $A_{V}$=0.0 to 6.0 in steps of 1.0.}
\label{colours} 
\end{figure*}

The dereddend colours in Table \ref{irmags} do not depend greatly on
the precise value of the assumed extinction. Decreasing the extinction
from $A_{V}$=5.6 to 5.0 changes the colours ($I-K$), $(J-H)$ and
$(H-K)$ by 0.17, 0.06 and 0.03 mag respectively. To compare the
dereddend infrared colours of RX J1914+24 with late-type dwarf stars
we show in Fig.\ref{colours} the colours of RX J1914+24 in the
($I-K$), $(J-H)$ and ($I-K$), $(H-K)$ planes together with old disk
dwarfs with spectral types K7 to M7.5 (Bessell 1991). It is clear that
the dereddened infrared colours of RX J1914+24 are not consistent with
a dwarf late main sequence secondary star. (We should add a note of
caution at this stage that our $I$ band observations were not
simultaneous with our $JHK$ observations, but were earlier by a week).

\begin{figure}
\begin{center}
\setlength{\unitlength}{1cm}
\begin{picture}(6,5)
\put(-3.5,-13.3){\includegraphics{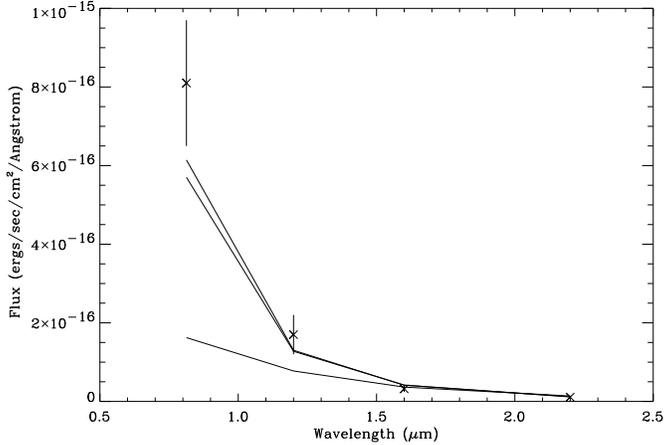}}
\end{picture}
\end{center}
\caption{The dereddened fluxes and three fixed temperature blackbody
fits. These correspond to blackbody temperatures (from top to bottom)
of 500000K, 50000K and 5000K respectively.}
\label{pasi}
\end{figure}

We now use the dereddend fluxes shown in Table \ref{irmags} to
estimate whether the spectral continuum could be fitted with a simple
blackbody model as would be roughly the case for a low temperature
white dwarf.  The errors in our data increase rapidly from $K$ towards
$I$, since they are dominated by the uncertainty in the amount of
reddening involved (the errors shown in Fig. \ref{pasi} include the
uncertainty in the photometric mean magnitude and also an uncertainty
in the extinction of $A_{V}=$0.5). Despite this, it is clear from our
blackbody fits, plotted in Fig. \ref{pasi}, that the spectrum is far
too hot to be explained simply by the blackbody component alone. In
Fig. \ref{pasi} we show the data and three fixed temperature blackbody
fits. No satisfactory blackbody fit was found to fit the data well,
although a higher temperature fit is less poor than a cooler one. If
the blackbody temperature is left as a free parameter in the fit, it
will tend towards infinity. Were we to assume that the extinction
towards RX J1914+24 was lower than $A_{V}=$5.6, (say $A_{V}=$4.0) then
we can obtain a good fit and lower temperature ($\sim$13000K).

Taking the extinction found by fitting the X-ray spectra as
representative of the optical extinction, these results suggest that
in the optical-IR region of the spectrum there is a component
additional to a blackbody, which adds to the emission at shorter
wavelengths. This could be due to emission from the accretion stream,
cyclotron emission from the white dwarf or an irradiated component on
the donor white dwarf. We now address these possibilities in turn.

Although we cannot exclude emission from a hot accretion stream this
seems unlikely because the $I$ band folded light curve
(Fig. \ref{fold}) would be double peaked unless we are viewing the
system at a low inclination. For a magnetic field of a few MG (see
below) the optically thin (and hence polarised) part of the cyclotron
spectrum is likely to be in red wavebands. Since we do not observe a
significant circularly polarised flux in $I$ the magnetic field is
either stronger than we expect (in which case the optically thin
cyclotron harmonics would be in the UV) or it is very low (in which
case the optically thin cyclotron harmonics would be in the far
IR). On the other hand it is also possible that a shock does not form
above the accretion region on the primary white dwarf because of the
small dimensions of the system.

Because the binary components are unusually close, we expect the
secondary white dwarf to intercept an appreciable part of the
accretion flux. Observations of conventional polars show that the
irradiated surface of the secondary is brighter in the trailing
hemisphere (eg Southwell et al 1995). Assuming that the irradiated
surface is much larger than the accretion region on the primary star,
then this would imply that the $I$ band flux peaks between
$\phi\sim$0.2--0.6 cycles after the X-ray peak -- which is indeed what
we observe (Fig. \ref{fold}). We conclude therefore that the variation
in $I$ is probably due to irradiation of the secondary by the X-rays
produced at the accretion region on the primary star.

If this is indeed the case, then we can also make the assumption that
the origin of the 9.5 min period seen in the $I$ band amplitude
spectrum (Fig. \ref{ipower}) is caused by the secondary star. On the
other hand the 9.5 min period seen in the {\sl ASCA} amplitude
spectrum (Fig. \ref{asca_amp}) is due to the accretion region on the
white dwarf. The fact that the amplitude peaks in the {\sl ASCA} and
$I$ band amplitude spectra are indistinguishable is further evidence
that RX J1914+24 is synchronised.

\section{The mass of the secondary star}
\label{masssec}

For semi-detached binaries, the Roche-lobe-filling condition implies a
unique relation between the average density of the donor star and the
binary orbital period (see eg Frank, King \& Raine 1992, eqn 4.8).
Assuming that the 9.5-min period detected in RX J1914+24 is indeed the
orbital period and that the Roche lobe approximation is valid for
degenerate stars, the average density of the secondary star will be
about 4600~g~cm$^{-3}$. The typical density of a main-sequence star is
$\sim 100$~g~cm$^{-3}$. Thus, the deduced high density of the
secondary star immediately places it into the class of degenerate
dwarfs. Using the Nauenberg (1972) mass-radius relation for white
dwarfs and the Eggleton (1983) approximation for the Roche-lobe volume
radius, we estimate that the mass of the secondary star is about
0.08~$M_\odot$.

\section{The distance to RX J1914+24}

If we assume for the sake of argument the 9.5-min seen in RX J1914+24
is the spin period and there is a second period (the orbital period)
which has been missed for some reason (ie the system is an IP) -- what
would the minimum distance be to the system?  Taking the empirical
relation for the spin and the orbital periods for IPs

\begin{equation}
 P_{\rm orb} \approx 10 P_{\rm spin}   
\end{equation} 

\noindent (Barrett, O'Donoghue \& Warner 1988), we estimate that the
orbital period would be $\approx$ 95~min. If the secondary star is a
main-sequence star filling its Roche lobe, a 95-min orbital period
implies a mass and radius of 0.12~$M_\odot$ and 0.16~$R_{\odot}$
respectively (eg Warner 1995).

Bailey (1981) derived a surface brightness relationship 
\begin{equation} 
 S_K = K + 5 - 5\ \log D+ 5\  \log r_2 \ , 
\end{equation}
 from which we can deduce the distance if the surface brightness of
the star in the $K$ band ($S_K$), the distance ($D$, in pc) and the
radius of the secondary star ($r_2$, in solar units) are known. The
mass and radius of the secondary star in RX J1914+24, deduced from the
assumed 95-min orbital period, implies a spectral type of K7 -- M3 for
the star. We have already noted that this is not consistent with our
IR colours. For main sequence stars, $S_{K}\sim 4.4$ (Ramseyer
1994). Using the de-reddend $K$ magnitude in table \ref{irmags}, we
obtain a value of 310~pc for the lower limit to distance of RX
J1914+24. This distance is consistent with the mean distance of known
CVs. However, if RX J1914+24 is at a distance of 310~pc, the
bolometric luminosity given in Cropper et\,al.\ (1998) should be
increased by a factor of $\sim$10. Then the mass transfer rate of the
system is now as high as $6 \times 10^{-6}~M_\odot/{\rm yr}$.  Such a
high mass transfer rate is difficult to be explained by current
orbital evolution theories. We therefore consider this as further
evidence for a double degenerate interpretation. If we have a longer
orbital period, ie $P_{\rm orb} > 10 P_{\rm spin}$, then this problem
is significantly worse.

We now go on to determine a distance based on the measurement of the
extinction to RX J1914+24. The solar system lies in a region of space
which has a lower hydrogen density compared to the surrounding region
(eg Warwick et al 1993).  The distance to the edge of this low density
region depends on the direction. This has been determined, for
instance, by the distribution in the sky of white dwarfs visible in
the soft X-ray band (eg Barstow et al 1997). Because the X-ray
spectrum of RX J1914+24 is heavily reddend (\S \ref{asca}), this
implies that it is at a distance greater than the distance to the edge
of the low density region. Using the map of distance to the edge of
the low density region as a function of galactic co-ordinates in
Barstow et al (1997) we find that the distance to RX J1914+24 is
greater than $\sim$100pc, assuming that the absorption we observe is
not internal to the binary system.

If we assume that the $I$ band flux can be characterised by a
blackbody, we find that a blackbody with the volume of a secondary
Roche lobe for a 9.5 min period will give the dereddend $I$ band flux
(Table \ref{irmags}) at a distance of 100pc if it has a temperature of
$\sim$9500K. For a distance of 400pc we need to increase the
temperature to $\sim$60000K to match the dereddend $I$ band flux.
While caution should be exercised when comparing the temperatures of
accreting white dwarfs with non accreting white dwarfs (since the
latter have no external heat source), non accreting white dwarfs
generally have temperatures \ltae 60000K (Barstow et al 1993). Whilst
there is some uncertainty regarding the appropriateness of adopting a
blackbody, it nevertheless suggests that RX J1914+24 lies at a
distance of $\sim$100--400pc. If there is significant heating of the
secondary in the form of irradiation, then the upper limit will be
less than 400pc.

\section{Long term variations in intensity}

\begin{figure*}
\begin{center}
\setlength{\unitlength}{1cm}
\begin{picture}(13,5)
\put(-1,-33){\includegraphics{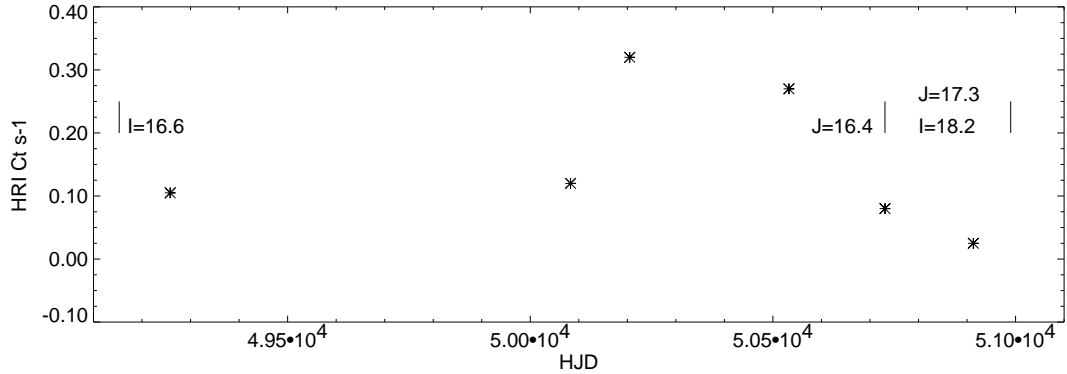}}
\end{picture}
\end{center}
\caption{The equivalent {\sl ROSAT} HRI peak count rate as a function
of time (HJD-2400000.0).}
\label{hri_time} 
\end{figure*}

In Fig. \ref{fold} we show the X-ray folded light curves taken at
various epochs over a space of 5 years. To compare the peak count rate
using {\sl ROSAT} PSPC and {\sl ASCA} to that of the {\sl ROSAT} HRI
we used the best fit spectral parameters of fitting the {\sl ASCA}
spectrum with an absorbed blackbody (\S \ref{asca}). The equivalent
HRI count rate as a function of time is shown in Fig.
\ref{hri_time}. We find that the X-ray flux varies by approximately an
order of magnitude. If this X-ray variation is reflected in the
optical-IR flux then this would correspond to a range in optical
brightness of 2.5 magnitudes. This is similar to the range seen in the
long term optical light curve of the polar AM Her (Feigelson, Dexter
\& Liller 1978). On the other hand, IPs (which have a reservoir of
material in the form of a disk), do not (in general) show such a large
range in brightness.  We also show the $I$ and $K$ band magnitudes at
two epochs which are broadly consistent with the change in the X-ray
flux. There is no evidence for a correlation between the shape of the
light curve (Fig. \ref{fold}) and the intensity.

The most obvious cause for the long term intensity changes is a
variation in the mass transfer rate. In conventional polars various
schemes have been proposed to explain these variations although none
have been fully developed.  One such scheme is that of magnetic
activity, such as star spots, near the $L_{1}$ point on the secondary
somehow affects \Mdot (eg Barrett, O'Donoghue \& Warner 1988, Livio \&
Pringle 1994).

\section{Discussion}

Cropper et al. (1998) argued that RX J1914+24 was a double degenerate
polar to account for the lack of a second period in the {\sl ROSAT}
data and the shape of the X-ray light curve. This is supported by our
$I$ band and {\sl ASCA} light curves and by the lack of any variation
at any other period in the $J$ band. Further the long term X-ray light
curve and the {\sl ASCA} spectrum is typical of polars.  These
observations reinforce the proposal of Cropper et al. (1998) that
RX J1914+24 is a double degenerate polar. We now go on to discuss
system parameters such as the primary stars magnetic field strength
and the forces which would be needed to keep RX J1914+24 synchronised.

\subsection{Lower limit to the primary's magnetic field} 
\label{magfield}

As no other periods apart from the 9.5 min period have been detected
in either the X-ray or $I$ band light curves, we conclude that RX
J1914+24 does not have an accretion disk. For accreting white
dwarfs, the absence of an accretion disk can be due to a strong white
dwarf magnetic field. When the Alfv\'en speed, which depends on the
local magnetic field strength and the density of the accretion flow is
much larger than the Keplerian speed at the circularisation radius, an
accretion disk cannot be formed. As an approximation, we may use the
criterion:
\begin{eqnarray}  
\label{magmom}
  \mu_1 & > &  1.7\times 10^{33} {\rm G~cm}^3
    \biggl({\eta \over 0.3}\biggr)  \nonumber \\ 
        &    & \times 
     \biggl( {{P_{\rm orb}}\over {60~{\rm min}}} \biggr)^{7/6} 
     \biggl( {{M_1} \over {M_\odot}}\biggr)^{5/6} 
     \biggl( {{\dot M} \over {10^{18}~{\rm g~s}^{-1}}}\biggr)^{1/2}    
\end{eqnarray} 
(Wickramasinghe, Wu \& Ferrario 1991) to deduce the lower limit to the
magnetic moment of the primary star, $\mu_1$, and hence the
corresponding limit to the white dwarf polar magnetic field strength.
Here, ${\dot M}$ is the accretion rate and $\eta$, which approximately
equals 0.3, is a parameter which takes into account the angle of the
stream as it leaves the inner Lagrangian point (Lubow \& Shu 1975).

\begin{figure}
\begin{center}
\setlength{\unitlength}{1cm}
\begin{picture}(6,7)
\put(-14.8,-2){\includegraphics{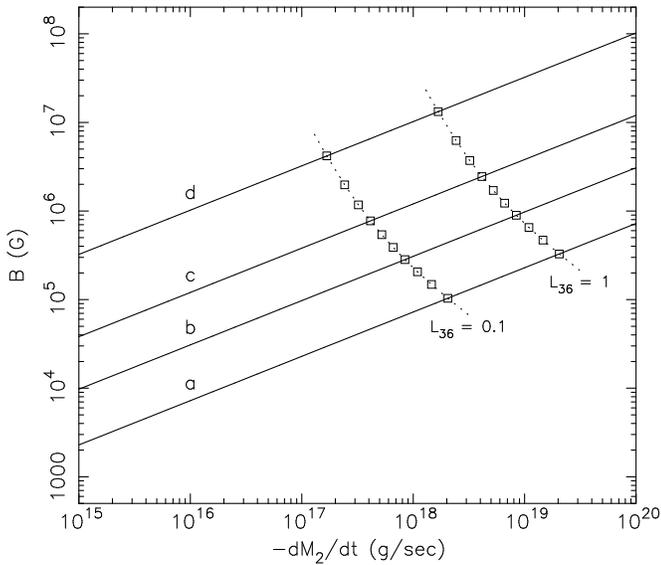}}
\end{picture}
\end{center}
\caption{The solid lines a, b, c and d show the relationship between
the polar magnetic field strength of the primary white dwarf and the
accretion rate such that the Alfv\'en speed and the Keplerian speed
are equal at the circularisation radius for 0.4, 0.7, 1.0 and 1.3\Msun
white dwarfs respectively. These provide a lower limit to the field
for diskless accretion. The binary orbital period is fixed at
9.5~min. The open squares show the lower limits to the polar magnetic
field strength with masses from 0.4 to 1.3\Msun (from bottom to top
with increments of 0.1\Msun), for accretion luminosities $L_{36}
(=L/10^{36}{\rm erg~s}^{-1}) = 0.1$ and 1.}
\label{kinwah}
\end{figure}

Using equation \ref{magmom} we show in Fig. \ref{kinwah} the surface
field strength required to equate the Alfv\'en speed and the Keplerian
speed at the circularisation radius, as a function of the accretion
rate.  We also show the lower limit to the polar magnetic field of the
primary white dwarf in binaries for various white dwarf masses $M_1$
and accretion luminosities $L$. The orbital period of the binary is
fixed at 9.5~min. As shown, if the accreting white dwarf is not
massive (\ltae~1$M_\odot$ and accretion rate is low
($<10^{18}$~g~s$^{-1}$)), a small polar field (\ltae\ 1~MG) is
sufficient to prevent the accretion disk formation.

From the {\sl ROSAT} PSPC data, Cropper et\,al. (1998) deduced that
the bolometric accretion luminosity of RX J1914+24 is $\sim10^{35} -
10^{36}$~erg~s$^{-1}$ (for a distance of 100~pc and hydrogen column
density $N_{_{\rm H}} \approx 1 \times 10^{22}~{\rm cm}^{-2}$).  If
the mass of the primary white dwarf in RX J1914+24 is about
1.0~$M_\odot$, the surface polar field strength of the white dwarf is
probably \gtae\ 1MG.
  
\subsection{The synchronising torques}  

While it is possible that an orbital period longer than the 9.5 min
period is hidden in the data, let us assume that RX J1914+24 is indeed
synchronised. The geometrical configuration and synchronism of
conventional polars have been investigated by many authors (eg
Chanmugam \& Wagner 1977; Joss, Katz \& Rappaport 1979; Lamb et\,al.\
1983; Campbell 1983, 1997; Lamb \& Melia 1987; King, Frank \&
Whitehurst 1990; Wu \& Wickramasinghe 1993). Some earlier studies (eg
Chanmugam \& Wagner 1977) implicitly implied that the diskless and
synchronisation conditions are equivalent. Although synchronism
ensures diskless accretion, diskless accretion may not imply perfect
synchronism.

A proper treatment of synchronisation involves magneto-hydrodynamic
(MHD) interaction between the stars and the plasma between them (see
Campbell 1997) and also the physics of stellar structures of magnetic
stars. The issue of synchronisation and formation of ultra-short
period double-degenerate binaries is beyond the scope of this paper
and deserves a separate study. Here we discuss synchronism in RX
J1914+24 briefly only in terms of a crude comparison between the
strengths of the accretion torque and the restoring magnetic torques.

The accretion torque which operates about the spin axis of an 
accreting star is given by 
\begin{eqnarray} 
\label{acctorque}
   N_{_{\rm acc}} & = & {{2 \pi \dot M R^2_{_{\rm L}}}
    \over P_{\rm orb}}   \nonumber \\ 
       {\ } &  \approx & 2 \times 10^{35}  ~{\rm dyne~cm}   
    \nonumber \\ 
  {\ }   &  {\ } &  \times
      \biggl({{\dot M} \over {{10^{18}~{\rm g/s}}}} \biggr)    
      \biggl({{P_{\rm orb}} \over {10~{\rm min}}} \biggr)^{1/3} 
      \biggl({M_1 \over {1.0~M_\odot}} \biggr)^{2/3} \,   
\end{eqnarray}  
where $P_{\rm orb}$ is the orbital period and $R_{_{\rm L}}$ is the
Roche lobe of the star. For a system with an accretion luminosity of
about $10^{36}$~erg~s$^{-1}$, which corresponds to a mass transfer
rate of $\sim 10^{18}$~g~s$^{-1}$, the accretion torque is $\sim
10^{35}~{\rm dyne~cm}$. Taking the accretion luminosities deduced in
Cropper et\,al. (1998), we find that the accretion torque on the
primary white dwarf in RX J1914+24 is 10 -- 100 times that of
conventional polars (see Wu \& Wickramasinghe 1993). We now
investigate whether this exceeds the available synchronising torque. 

In conventional polars, the accretion torque is counterbalanced by the
magnetic torque and the system maintains synchronism. Two types of
magnetic restoring torque have been proposed. The first type is due
to the torque arising from magneto-static dipole/multipole interaction
between the primary and the secondary star. The second type is due to
the torque arising from the MHD interaction between the stars and the
plasmas between the stars. In general the second type of torque
requires that the secondary star, which is a main-sequence star, has a
convective envelope threaded by the magnetic field of the primary
degenerate star.  Ohmic dissipation occurs at the envelope of the
secondary star when the system departs from synchronism.

We can make an estimate of the torque due to magneto-static
interaction between the dipole/multipole components of the primary
star and the dipole of the secondary star using the equations in Wu \&
Wickramasinghe (1993). For an orbital period of 9.5 min, the orbital
separation is $\sim10^{10}$ cm. The ratio of orbital separation to the
radius of the accreting white dwarf is $\sim$20 for a 1\Msun white
dwarf. For $B$=10MG and $\mu_{1}=1-10\times10^{34}$ G cm$^{3}$ we find
from Wu \& Wickramasinghe (1993) that the torque due to magneto-static
interaction is of the order of 10$^{35}$ dyne cm or less. (There is
some uncertainty since we do not know the magnetic field
configuration). This is of the same order or less than the accretion
torque (equation \ref{acctorque}). Therefore it is not clear if this
mechanism alone is able to maintain the synchronism of RX J1914+24.

Degenerate white dwarfs are highly conducting bodies. When synchronism
is disturbed, ohmic dissipation occurs much more efficiently in the
plasma between the stars than in the stellar envelopes of the white
dwarfs. Thus, the large scale MHD instabilities of the field
configuration will determine the strength of the restoring magnetic
torque (Lamb et\,al. 1983).  According to Lamb \& Melia (1988), if
both the primary and secondary stars have a non-negligible magnetic
moments, the MHD torque acting on the primary star is
\begin{eqnarray}
\label{mhd} 
 N_{_{\rm MHD}} & \approx & 5\times  10^{35}~{\rm dyne~cm}  
  \biggl( {\gamma \over 0.5} \biggr)
  \biggl({a \over {{10^{10}~{\rm cm}}}} \biggr)^{-3} 
      \nonumber \\ 
     &   &   {\ } \times 
      \biggl({{\mu_1} \over {{10^{33}~{\rm G~cm}^3}}} \biggr)  
      \biggl({{\mu_2} \over {{10^{33}~{\rm G~cm}^3}}} \biggr)    
\end{eqnarray} 
where $\mu_1$ and $\mu_2$ are the magnetic moments of the primary and
the companion stars respectively, $\gamma$ is the pitch angle of the
dipole magnetic field of the primary star at the companion star, and
$a$ is the orbital separation. For the configurations of polars, the
value of $\gamma$ is approximately 1 (Low 1982; Lamb et\,al.\ 1983).
   
For a primary with mass $\sim$1\Msun, Fig. \ref{kinwah} implies that
the surface magnetic field of the primary is \gtae\ a few MG if we
assume a mass transfer rate similar to that observed by Cropper et
al. (1998). This implies a magnetic moment $\mu_1 \sim 10^{34}~{\rm
G~cm}^3$. Thus, provided that the secondary star has a similar
intrinsic or induced magnetic moment, Equation \ref{mhd} implies that
the MHD synchronising torque is of the order of $5\times10^{37}$ dyne
cm. If the mass transfer rate is not much higher than
$10^{18}$~g~s$^{-1}$, the MHD synchronising torque is several order of
magnitudes greater than the accreting torque (equation
\ref{acctorque}) and is able to ensure that the system is in
synchronous rotation.

It is worth noting that in deriving equation \ref{acctorque} the
magnetic field is ignored and the angular momentum is assumed to be
transferred from the accretion flow to the star directly. For a
magnetically channeled flow, the interaction between the accretion
flow and the star is more complicated. A recent study (Li,
Wickramasinghe \& R\"udiger 1996) has shown that angular momentum can
be transferred from the accretion flow to the star only via the
magnetic stress. In non-synchronously rotating systems, the accretion
flow interacts with the magnetic field of the accreting star. In
polars, as the field lines of the two stars are interconnected, the
accretion flow interacts magnetically with both stars. This may allow
the angular momentum of the accretion flow to distribute between the
two stars and hence to be transferred to the orbit. Thus, equation
\ref{acctorque} probably overestimates the strength of these first
order considerations of the destabilising torque due to accretion. The
synchronism of RX J1914+24 is therefore likely to be more stable than
expected from a balance between the MHD torque and the conventional
accretion torque.
 
\subsection{Mass transfer rate} 

The mass transfer in low mass semi-detached binaries is generally
driven by the orbital evolution. When the orbital angular momentum of
a binary decreases, the orbital separation shrinks, causing matter to
overflow the Roche lobe of the secondary star and accrete onto the
primary star. For short (\ltae~3~hr) period systems, gravitational
radiation is an efficient process for angular momentum loss, while for
long (\gtae~3~hr) period systems magnetic braking is the dominant
process (eg Verbunt \& Zwaan 1981).

Magnetic braking requires the presence of a stellar wind from the
secondary star by which the angular momentum is transported. In
double-degenerate systems there is no stellar wind. Magnetic braking
therefore cannot operate, regardless of the orbital period and whether
the system is synchronised or not (see Li, Wu \& Wickramasinghe 1994).
If RX J1914+24 is indeed a double degenerate system, then the orbital
evolution of RX J1914+24 should be solely driven by gravitational
radiation.

The mass transfer rate for gravitational-radiation driven orbital
evolution is:
\begin{eqnarray}     
   \dot M  & \approx & 1.5 \times 10^{-6}~M_\odot/{\rm yr} 
    \nonumber \\ 
    &   & \times \biggl[ {{m_1^2 m_2^2} \over {(\alpha m_1-m_2)}}  
         {1 \over {(m_1+m_2)^{1/3}}} \biggr] 
    \biggl({P_{\rm orb} \over {10~{\rm min}}}\biggr)^{-8/3}\ ,  
\end{eqnarray} 
(cf Faulkner 1971; Wickramasinghe \& Wu 1994), where $m_1 =
M_1/M_\odot$, $m_2 = M_2/M_\odot$ and $\alpha$ = 5/6 + $n$/2, where
$n$ is related by the form $R_{2} \propto M_{2}^{n}$. For normal
stars, $n\sim$1, giving $\alpha$=4/3, while for low mass He white
dwarfs, $n\sim$ --1/3, giving $\alpha\sim$2/3 (Nauenberg 1972). For
$m_1 = 1.0$, $m_2 = 0.08$ (cf \S \ref{masssec}) and an orbital period
of 9.5 mins, the expected mass transfer rate is $\dot M \approx 1.8
\times 10^{-8}~M_\odot~{\rm yr}^{-1} \approx 1.1 \times 10^{18}~{\rm
g~s}^{-1}$ for a He white dwarf. This is consistent with the value
(Cropper et\, al.\ 1998) deduced from the ROSAT PSPC data. As mass
transfer in RX J1914+24 cannot be driven by magnetic braking or
nuclear evolution, the agreement between the observed accretion
luminosity and predicted mass transfer rate show evidence of the
operation of gravitational radiation in close binaries.

\section{Conclusions}

We have found the optical counterpart of RX J1914+24. Our new optical
and X-ray data strengthen the suggestion of Cropper et al (1998) that
RX J1914+24 is the first known double degenerate polar and has the
shortest orbital period of any known binary system. We suggest that
the variation in the $I$ band flux is due to the irradiated face of
the secondary white dwarf. The long term X-ray light curve shows a
variation in flux of an order of magnitude (typical of polars). We
estimate that the secondary white dwarf has very low mass (0.08\Msun)
and the primary white dwarf has a magnetic field strength greater than
a few MG. For double degenerate binaries the dominating synchronising
torque is that of MHD interaction between the two stars and the
accretion stream while mass transfer is driven solely by gravitational
radiation.

\section{Acknowledgments}

We would like to thank both the staff of the NOT and the UKIRT for
their help at the telescopes. Some of the data presented in this paper
have been taken using ALFOSC, which is owned by the Instituto
deAstrofisica de Andalucia (IAA) and operated at the Nordic Optical
Telescope under agreement between IAA and the NBIfA of the
Astronomical Observatory of Copenhagen. UKIRT is operated by the Joint
Astronomy Centre on behalf of PPARC. We gratefully acknowledge Darragh
O'Donoghue for the use of his period analysis software. KW
acknowledges the support from the ARC through an Australian Research
Fellowship and PH acknowledges the support from the Academy of Finland.

\end{document}